\documentstyle[epsf]{mn}
\begin{document}
\newcommand{\twiddles}{\sim}
\newcommand{\etal}{{et al}\/.}
\newcommand{\uv}{{\it uv}}
\def\Ssin#1C#2 {#1C\,#2}
\def\Ss#1{\Ssin#1 }
\def\Ssf#1{\Ssin#1 }
\def\Ssq#1{}
\def\afterpage#1{}
\def\dgr{^\circ}
\title{The jets in \Ss{3C296}}
\author[M.J.~Hardcastle \etal]{M.J.~Hardcastle\thanks{Present address:
H.H.\ Wills Physics Laboratory, University of Bristol, Royal Fort,
Tyndall Avenue, Bristol BS8 1TL. E-mail: M.Hardcastle@bristol.ac.uk.}, P.~Alexander,
G.G.~Pooley and J.M.~Riley\\Mullard Radio Astronomy Observatory, Cavendish
Laboratory, Madingley Road, Cambridge, CB3 0HE}
\maketitle
\begin{abstract}
We present observations made with the VLA at 1.5 and 8.4
GHz of the nearby FRI radio galaxy \Ss{3C296}. The most recent models
of FRI radio galaxies suggest that substantial deceleration must take
place in their jets, with strongly relativistic velocities on parsec
scales giving place to at most mildly relativistic velocities on
scales of tens of kiloparsecs. The region over which this deceleration
takes place is therefore of considerable interest. By considering the
side-to-side asymmetries of the jets of \Ss{3C296}, we constrain the
region of strong deceleration in the source. Our observations show
evidence that the jets have slow edges surrounding faster central
spines. We discuss the implications of our observations for models of
the magnetic field structure in these objects.
\end{abstract}

\section{Introduction}

FRI radio galaxies (Fanaroff \& Riley 1974) very often show a pair of
bright jets emanating from a central core; the jets may eventually
form lobes or just disappear into the noise, as in the `archetypal'
twin-jet FRI, \Ss{3C31} (Fomalont \etal\ 1980; Laing 1996). It is well
known (Burch 1979) that the brightness decline in such jets is
sub-adiabatic; that is, the jet surface brightness drops much less
rapidly with distance along the jet than would be expected for a
smooth uniformly expanding jet with constant non-relativistic velocity
and no particle acceleration. One plausible explanation for this is
that the boundary layers of the jets are turbulent and entrain
external material; this would both cause them to decelerate and allow
for some particle acceleration, as bulk jet velocity is converted into
radiative particle velocity by a second-order Fermi process. Fits of
such models have been made to jets in large samples of FRIs (Bicknell
\etal\ 1990) suggesting velocities in the jets of at most a few
$\times 10^4$ km s$^{-1}$.

At the same time, images on kiloparsec- or hundred-parsec scales of
these jets very often show that one jet (often just referred to as
`the jet') is very much brighter than the other (`the counterjet')
near the radio core. The effect lessens with distance from the core,
but may persist for many kiloparsecs. VLBI imaging of objects with
this sort of asymmetry shows a one-sided parsec-scale jet pointing
towards the brighter kiloparsec-scale jet (e.g.\ Giovannini \etal\
1993; Pearson 1996). The most popular explanation for this type of
behaviour is beaming due to relativistic bulk velocities of the
emitting material; if this is the case, both the parsec- and
kiloparsec-scale jets must be relativistic and there must be
substantial deceleration between scales of a few kiloparsec (where one
jet is much brighter than the other) and a few tens of kiloparsecs
(where the jets are of similar surface brightness). Bicknell (1994),
among others, has analysed the dynamics of a decelerating relativistic
jet and shown that this sort of deceleration is feasible. However, it
remains to be seen whether the details of this sort of model are
consistent with observation.

Detailed, high-resolution images of this type of object are
surprisingly rare. In an earlier paper (Hardcastle \etal\ 1996) we
presented images of a twin-jet FRI radio galaxy, \Ss{3C66B}, and
showed that the brightness asymmetries in its inner jets were
consistent with the relativistic beaming model and with the velocities
predicted by the model of Bicknell (1994), within the limits imposed
by the curved jet and counterjet in that source. We found that the
apparent magnetic field (the emission-weighted projection onto the
plane of the sky of the true three-dimensional field structure) was
parallel to the axis of the jet in the inner 4 kpc on the jet side,
turning further out to a transverse configuration with a
longitudinal-field sheath.  We discussed the evidence for this sort of
behaviour in other FRIs. Recently Laing (1996) has presented images of
another twin-jet object, \Ss{3C31}, which shows similar magnetic field
behaviour. The straight twin jets in \Ss{3C31} allowed him to produce
a `sidedness map' of the object, showing the ratio between the surface
brightnesses of jet and counterjet as a function of position in the
jet; the jet-counterjet ratio was shown to be greatest on the jet
ridge line, and Laing discussed this in terms of a model in which the
jets in these objects initially consist of a fast `spine' with a
relativistic velocity and a `shear layer', containing material with a
range of velocities between that of the spine and zero, in which the
interaction with the external medium is presumed to take place.

\Ss{3C296} is another nearby bright FRI radio galaxy, and is an
excellent object to study in the context of relativistic beaming
models, because its jets (like those of \Ss{3C31}) are straight. In
this paper we present detailed observations of the total intensity and
polarization of its jets, and discuss the extent to which it fits
proposed models for relativistic beaming and deceleration.

\Ss{3C296} has a redshift of 0.0237 ($P_{178} = 2.8 \times 10^{24}$ W
Hz$^{-1}$ sr$^{-1}$; at this distance 1 arcsec = 0.67 kpc, and the
total linear size of the radio source is 300 kpc). In the radio it was
imaged by Leahy \& Perley (1991, hereafter LP); earlier maps are in
Leahy (1985) and Birkinshaw, Laing \& Peacock (1981). The host galaxy
(\Ss{NGC5532}) is a large elliptical (Owen \& Laing 1989). It is
located in the Abell cluster A1890 and has a close small companion
(Wyndham 1966; Colina \& P\'erez-Fournon 1990). Fabbiano \etal\ (1984)
report a detection with EINSTEIN. There are no published VLBI
observations.

Throughout this paper we use a cosmology in which $H_0 = 50{\rm\
km\,s^{-1}\,Mpc^{-1}}$ and $q_0 = 0$.

\section{Observations}

We observed \Ss{3C296} with all four configurations of the VLA, with
dates and integration times as shown in table \ref{obs}. The 1.5-GHz
data from B, C and D configurations were kindly supplied by
J.P.~Leahy; details of these observations may be found in LP.

\Ss{3C286} was used as a primary flux calibrator and polarization
angle calibrator for all our new observations. The phase calibrator
was the nearby point source 1413+135, observed at intervals of around
30 min; this object normally doubled as a point source for
polarization calibration, but in cases where the observations were
interleaved with integrations on other objects other bright point
sources were occasionally used.

The data were reduced within {\sc aips} in the standard way. The datasets
from each configuration were initially reduced separately, each
undergoing several iterations of CLEANing and phase self-calibration.
For the 8.4-GHz observations, the B, C and D-configuration datasets
were then phase-calibrated, using the appropriate baselines, with
images made from the higher-resolution datasets, with which they were
then merged without reweighting. Thus the B-configuration data were phase
calibrated with an image made from the A-configuration data and merged
with it to form an AB dataset; images made with this at low resolution
were used to phase calibrate the C-configuration data and the two were
merged to form an ABC dataset, and so on. This process ensures phase
consistency in the data while removing the need for a self-calibration
of the final merged dataset. For the 1.5-GHz observations, the
A-configuration dataset, after self-calibration, was concatenated
without reweighting with the existing BCD dataset and the combined
dataset self-calibrated.  This process has the potential to introduce
errors into the short baselines of the resulting dataset, but in this
case any such errors were insignificant.

Maps were made using the {\sc aips} task IMAGR, with tapering of the
$u$-$v$ plane where low-resolution maps were required. The robustness
parameter in IMAGR was used to temper the uniform weighting of the
$u$-$v$ plane, to improve the signal-to-noise ratio. To make the
full-resolution map, we used a robustness of $-1.5$ and no tapering,
as experiment showed that this gave the best trade-off between
resolution and sensitivity; the lower-resolution maps were made with
robustness $0$ and tapering chosen to give a good match between the
required beam size and the fit to the dirty beam made by IMAGR. In all
cases the restoring beam was a circularly symmetrical Gaussian and the
resolution quoted is its FWHM.

The 8.4-GHz dataset seriously undersamples the large-scale structure
of the source, which has a largest angular size of approximately 440
arcsec: the largest structure which can be reproduced with the VLA at
8.4 GHz with a single pointing is around 180 arcsec. However, the
missing baselines are unlikely to affect imaging of the jets, and our
high-resolution images show no significant indications of
undersampling, such as a negative `bowl' around the source. Because
our experience has been that undersampling of this kind can cause the
{\sc aips} task VTESS to converge poorly, and because the images are
dominated by the point-source core, we elected not to make a
maximum-entropy image.

Low-resolution 1.5-GHz images of the source are presented by LP, and
as our A-configuration observations add little information on these
scales, maps made from the combined 1.5-GHz dataset are not presented here.
In considering the structures of the jets, it must be borne in mind
that they are superposed on large diffuse lobes, which are not seen in
our 8.4-GHz maps because of a combination of undersampling and low
radio surface brightness. This makes it difficult to define an
unambiguous jet edge.

\section{Results}

\subsection{The core}

The radio core of \Ss{3C296} did not vary over the timescales of our
observations either at 8.4 or 1.4 GHz, within the errors imposed by
the uncertainty of absolute flux calibration at the VLA. Its flux at
8.4 GHz was 126 mJy and at 1.4 GHz 58 mJy.

\subsection{Structure at high resolution}

\begin{figure*}
\begin{center}
\leavevmode
\vbox{\epsfysize 12cm\epsfbox{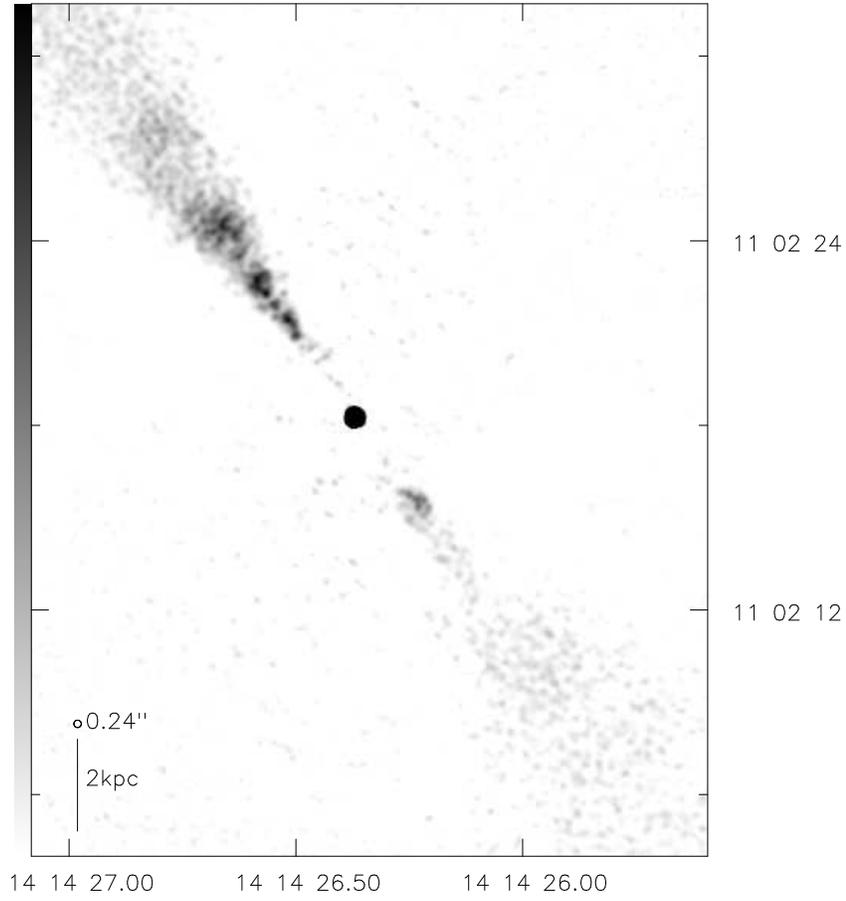}}
\caption{\Ssf{3C296}
at 8.4 GHz, 0.24-arcsec resolution. Linear greyscale; black is 0.35 mJy
beam$^{-1}$. R.m.s.\ off-source noise is 14 $\mu$Jy.}
\label{3C296.024}
\end{center}
\end{figure*}

In Fig.\ \ref{3C296.024} the inner parts of the jets of the source
can be seen. The jet brightens at approximately 2 kpc from the
nucleus. It appears to have a row of central compact knots in the
first 2 kiloparsecs of bright emission, and is increasingly diffuse
thereafter. There is little evidence for a bright FRII-like inner jet
as seen in some other objects (e.g.~\Ss{3C66B}, Hardcastle \etal\
1996). A knot in the inner part of the counterjet, again starting at
around 2 kpc from the core, is seen to be elongated, with some faint
compact structure, but there is no other compact structure in the
counterjet.

\subsection{Structure at intermediate resolution}

\begin{figure*}
\begin{center}
\leavevmode
\vbox{\epsfysize 12cm\epsfbox{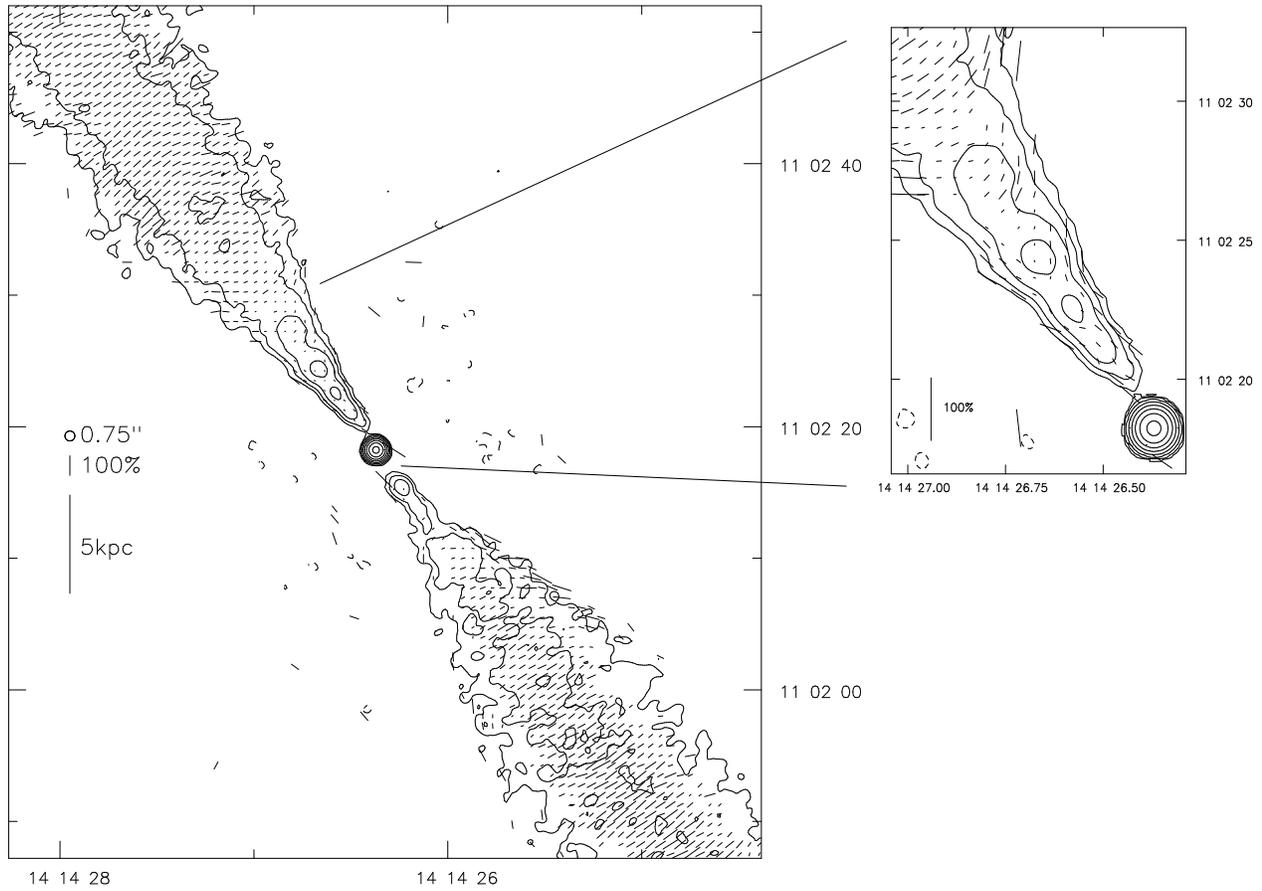}}
\caption{\Ssf{3C296}
at 8.4 GHz, 0.75-arcsec resolution. Contours at $0.2 \times
(-2\protect\sqrt 2, -1, 1, 2\protect\sqrt 2, 8, 16\protect\sqrt 2, \dots$) mJy
beam$^{-1}$. R.m.s.\ off-source noise is 11 $\mu$Jy. Vectors show
direction of $B$-field, and their length indicates degree of
polarization. Inset shows details of the polarization in the inner jet.}
\label{3C296.075}
\end{center}
\end{figure*}

Fig.\ \ref{3C296.075} shows the jets at 0.75-arcsec resolution. The
knot in the counterjet is notable again at this resolution; an almost
identical knot is seen in the maps of \Ss{3C31} by Laing (1996).  In
this image the counterjet appears marginally wider than the jet,
particularly at around 20 arcsec from the nucleus, but the difference
is very slight and is not visible on lower-resolution,
higher-sensitivity maps.

The polarization map includes a correction for Ricean bias and shows
all points with polarized and total intensity greater than three times
the respective off-source r.m.s.\ noise values. The position-angle
vectors are perpendicular to the observed $E$-field, and so show
magnetic field direction if Faraday rotation is negligible; that this
is so is suggested by the integrated rotation measure of -3 rad
m$^{-2}$ (Simard-Normandin, Kronberg \& Button 1981) and by the
rotation measure maps of Leahy (1985).

\begin{figure}
\vbox{\vskip 10pt
\epsfxsize \linewidth\epsfbox{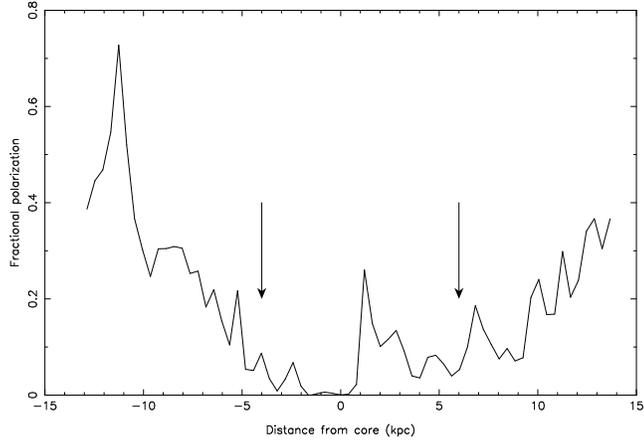}
}
\caption{Degree of polarization as a function of distance from the
core, measured from the 0.75-arcsec resolution map by averaging over
square elements (0.6 x 0.6 arcsec in size) along the ridge
line. Positive distances denote measurements in the jet, negative
distances denote the counterjet. Arrows mark the approximate locations
of the transition from longitudinal to transverse central apparent field.}
\label{3C296-fpol}
\end{figure}

The magnetic field is longitudinal for the first 6 kpc on the jet side
and transverse thereafter; the change in field direction is
accompanied by an increase in the degree of polarization from $\sim
10$ per cent to $\sim 20$ per cent. The polarization then rises to
$\sim 40$ per cent by 15 kpc out on the jet side. This can be seen on
a plot of degree of polarization on the ridge line against distance
(Fig.\ \ref{3C296-fpol}). The degree of polarization at the edges of
the jet is very low in the region where the central jet field changes
from a longitudinal to a transverse configuration. (This is also
observed in \Ss{3C31}.) On the counterjet side there is some
indication of a longitudinal magnetic field in the inner few
kiloparsecs (the L-band polarization maps, not shown, confirm that the
knot in the counterjet has longitudinal field), but the field is
definitely transverse to the jet direction by 4 kpc out. A
longitudinal-field sheath, with a high degree of polarization,
surrounds the counterjet (with a region of low polarization between
the transverse- and longitudinal-field r\'egimes), but there is no
indication of such a sheath on the jet side. This is also apparent in
the polarization maps of LP.

\subsection{Depolarization and spectral index}

There is no evidence for significant depolarization or rotation of the
magnetic field direction in the jet or counterjet from maps made with
matched baselines at 1.5 and 8.4 GHz. This is consistent with the low
integrated rotation measure mentioned above; it would appear that
there is little contribution from the local environment of \Ss{3C296}
to the Faraday depth towards the source on any scale. However, the
southern lobe is the more depolarized in low-frequency observations
(Garrington, Holmes \& Saikia 1996), which is in the sense expected if
relativistic beaming is responsible for the jet-counterjet
asymmetry. The spectral index between 1.5 and 8.4 GHz in the
inner part of the jet and counterjet is 0.6 (throughout spectral index
$\alpha$ is defined in the sense $S_{\nu} \propto \nu^{-\alpha}$), and
there is no evidence for spectral steepening along or across the jet
in the areas where both maps are reliable (out to about 50 arcsec),
consistent with the results of Birkinshaw \etal\ (1981).

\section{Discussion}

\subsection{Velocities in the jets}

If relativistic beaming affects the surface brightnesses of two
segments of an otherwise symmetrical pair of jets which both have a
bulk velocity $\beta c$ away from the core, then the observed
jet-counterjet ratio $R$ at that point is given by a standard formula
\[
R=\left({{1 + \beta \cos \theta}\over{1-\beta \cos
\theta}}\right)^{2+\alpha} \] 
where $\theta$ is the angle made by the jet to the line of sight.
From measurements of the jet-counterjet ratio we can measure the
component of the velocity along the line of sight, $\beta \cos
\theta$, and so constrain the velocity and $\theta$. It must be borne
in mind that the exponent $(2+\alpha)$ is in general correct only for a jet
which radiates isotropically in the rest frame, and that it may be
significantly in error when the magnetic field is fully ordered in a
direction with a component along the line of the bulk velocity
of the jet (Laing 1996: Cawthorne 1991). The low degree of
polarization in the jets, and the fact that the magnetic field is in
general transverse to them, imply that this effect can be neglected
in what follows. More seriously, if (as seems likely) the jets do not
have a single bulk velocity, then the jet-counterjet ratio at any
given point on a map gives us in some sense an emission-weighted
average of the velocities along the lines of sight through the jet and
counterjet.

\begin{figure*}
\begin{center}
\leavevmode
\epsfysize 10cm\epsfbox{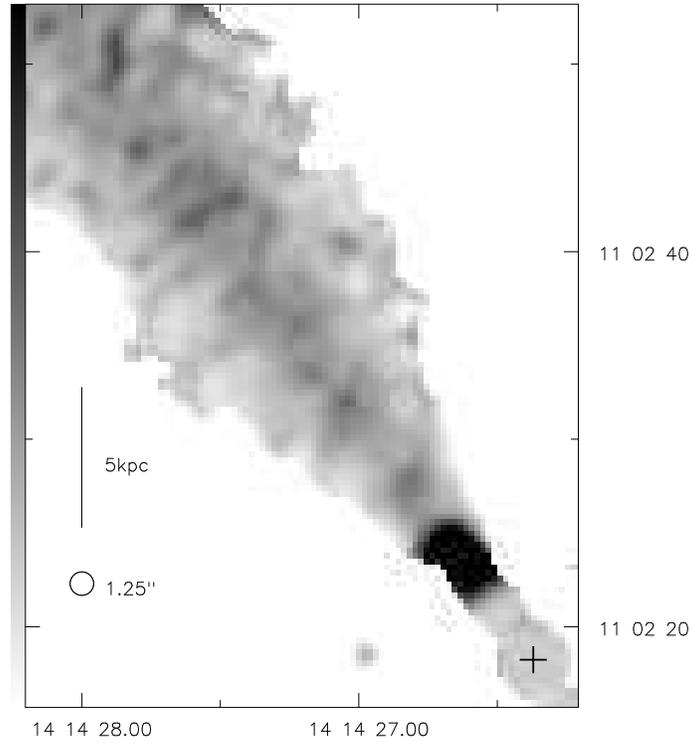}
\caption{Sidedness map of \Ssf{3C296}
at 8.4 GHz, 1.25-arcsec resolution, showing jet-counterjet ratio as a function
of position in the jet in the upper quarter of Fig.\ \ref{3C296.075}. Linear
greyscale: white is $R=0$ and black indicates $R > 5$. A cross
indicates the position of the core.}
\label{3C296-sided}
\end{center}
\end{figure*}

\begin{figure}
\vbox{\vskip 10pt
\epsfxsize \linewidth\epsfbox{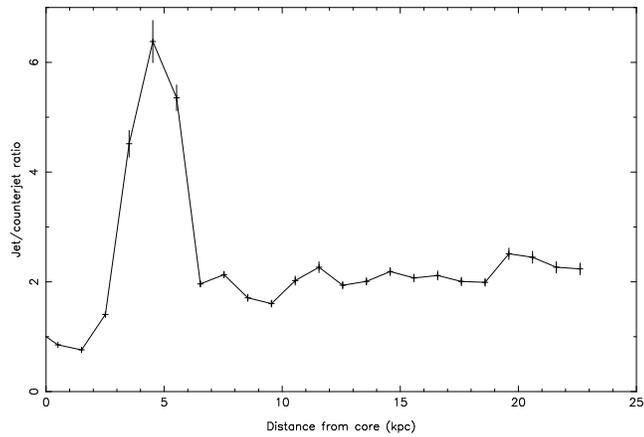}
}
\caption{Jet-counterjet ratio as a function of distance from the core
over the central 3 arcsec (in width) of the jet. The jet and
counterjet fluxes from 1.25-arcsec resolution maps were averaged in
bins 1.5 arcsec long and the ratio is plotted for each bin. Error bars
show the one-sigma error derived from the off-source noise.}
\label{3C296-sslice}
\end{figure}

\begin{figure}
\vbox{\vskip 10pt
\epsfxsize \linewidth\epsfbox{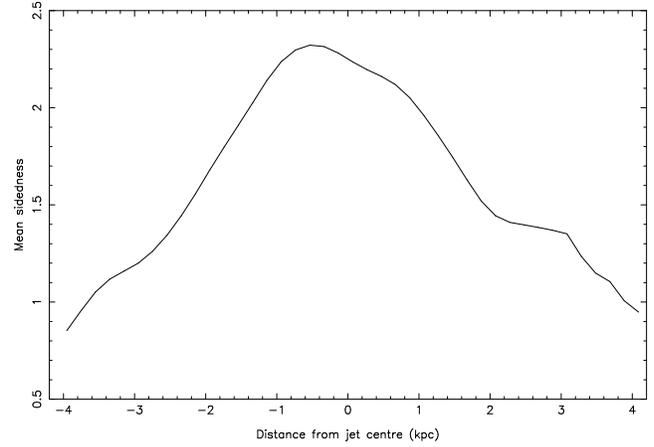}
}
\caption{Jet-counterjet ratio as a function of distance from the
mid-line of the jet, averaged in linear slices between 12 and 24 kpc
from the core.}
\label{3C296-sslice2}
\end{figure}

With this in mind, we follow Laing (1996) and produce a `sidedness
map' of the source, shown in Fig.\ \ref{3C296-sided}. This is made by
rotating an image, centred on the core, through $180\dgr$ and dividing
the original image by the rotated version; it thus shows
jet-counterjet ratio as a function of position in the jet. The map
shown here is made at 1.25-arcsec resolution in order to improve the
signal-to-noise ratio in the fainter parts of the jet, and only points
with a flux density greater than five times the r.m.s.\ off-source
noise in both rotated and original maps are used. The greatest
jet-counterjet ratio on this map (on the axis of the jet at $\sim 4$
kpc from the core) is 12, which translates using the formula above to
$\beta \cos \theta = 0.44$; thus $\theta \la 63\dgr$ in the beaming
model, and the velocity of the jet at its fastest is $\ga
0.44c$. Fig.\ \ref{3C296-sslice} shows the sidedness of the central
regions as a function of distance along the jet. It will be seen that
the jet is very asymmetrical only at the base, with the side-to-side
asymmetry parameter $R$ falling, by 6 kpc (9 arcsec) from the core, to
values around 2, corresponding to $\beta \cos \theta \approx 0.15$;
this drop in $R$ corresponds in position to the change from parallel
to perpendicular field and implies substantial deceleration at around
this point (the inferred velocity drops by a factor 2 on scales of 1
kpc, though this does not imply that the true deceleration is this
dramatic if there is a range of bulk velocities in the jet). There is
then a long region where the sidedness remains constant, and here
Fig.\ \ref{3C296-sided} shows that the central regions of the jet are
more one-sided than the edges, as found in \Ss{3C31} by Laing
(1996). The degree to which this is true is illustrated by a plot of
the mean sidedness as a function of distance across the jet, Fig.\
\ref{3C296-sslice2}; this was derived by taking the mean sidedness in
linear slices from Fig.\ \ref{3C296-sided} along the jet axis in the
region between 12 and 24 kpc from the core where the jet widths are
approximately constant. The mean value of the sidedness parameter in
the edges is close to 1; it is therefore possible that we are seeing
almost stationary material, but this low-surface-brightness region
will be most strongly affected by any departure from jet-counterjet
symmetry. Further out (at distances of $\sim 40$ kpc from the core)
the jet becomes more one-sided, but this coincides with a bend in the
ridge line of the jet and can be explained simply as brightening due
to compression; thereafter there is no significant difference in
surface brightness between the jet and counterjet.

The central regions of the jet are therefore brighter than the
corresponding regions of the counterjet over a projected distance of
some 30 kpc (corresponding, with our constraints on the angle to the
line of sight, to a true distance of $\ga 35$ kpc), and if we
attribute this to relativistic beaming then $\beta$ in the central
`spine' lies between approximately 0.15 and 0.35 over this whole
distance. These velocities are comparable to those found by Laing
(1996) in the equivalent region of \Ss{3C31}. We may conclude that the
model in which the central spine of the jet is moving
relativistically and substantially faster than its edges stands up
to this observational test.

\subsection{Polarization}

Two distinct features of the polarization of FRI radio sources must be
explained by models: these are the change from a longitudinal to a
transverse apparent field configuration in the inner few kpc of the
jet and the longitudinal-field `sheath' that is often seen around the
regions of the jet that have a transverse field. In the models
discussed by Laing (1993, 1996) these polarization features are
related to the velocity structure of the jet. He proposes that the
central spine always has a transverse magnetic field with no component
along the jet axis, and that the slower outer sheath of the jet has a
field that is either purely longitudinal or two-dimensional (either of
which will produce an apparent magnetic field direction parallel to
the jet when seen edge-on). A longitudinal or two-dimensional field is expected
in a region with strong shear (Begelman, Blandford \& Rees 1984). The
region of longitudinal apparent field close to the core is then
explained as emission from the shear layer, which dominates the
transverse-field spine in this region either because its degree of
polarization is intrinsically higher (as is the case for a sheath with
purely longitudinal field) and/or because of Doppler dimming of the
emission from the spine. Laing (1996) discusses the ways in which
two-dimensional and pure longitudinal fields can be distinguished
observationally, and it may be noted that the fact that the transition
from longitudinal to transverse apparent fields occurs at a greater
distance from the nucleus in the jet of \Ss{3C296} (as it does in
\Ss{3C31}) would imply a two-dimensional field in the shear layer.

An application of this model to \Ss{3C296} reveals two
problems. Firstly, we should expect the emission from the regions of
longitudinal apparent field near to the nucleus to be edge-brightened,
or at least only weakly centre-brightened, because of the assumed
Doppler dimming of the transverse-field spine. In fact, as the maps in
Figs \ref{3C296.024} and \ref{3C296.075} show, the inner few kpc of
the jet appear to be knotty and centre-brightened\footnote{We showed
in Hardcastle \etal\ (1996) that this problem is also implied by our
observations of the jets in \Ss{3C66B}.}. It is conceivable that the
knots seen in the central regions are projected filaments in the shear
layer, however.

Secondly, and perhaps more seriously, we should expect to see a
longitudinal apparent field in all emission which we can infer by
other means to be part of a shear layer; it is therefore surprising
that the edge of the jet in \Ss{3C296}, with jet-counterjet surface
brightness ratio between 2 and 1, should have an inferred magnetic
field that is clearly transverse to the jet (in contrast to the
counterjet, which has a longitudinal-field sheath in the equivalent
region). This contrasts with the excellent association found by Laing
(1996) in \Ss{3C31} between the edges of the jet with little apparent
beaming and the longitudinal-field sheaths. This effect cannot easily
be explained within a model with two entirely symmetrical jets. Some
environmental asymmetry is clearly present; this is indicated by the
different shapes of the lobes of the source on large scales and by the
slight bending of the jets at distances $>40$ kpc from the
nucleus. The environments of the edges of the jets, and thus the
entrainment rates, may therefore be different, and it is also hard to
decide whether either or both jets lie within the lobes rather than
being projected on to them. Nevertheless, in the relativistic beaming
model there must be significant shear in the jet at {\it some}
transverse distance from the ridge line, and there is no evidence for
the expected transverse field either in our maps or in the maps of LP.

\section{Conclusions}

The jet-counterjet surface brightness ratios in \Ss{3C296} are
consistent with recently developed models for decelerating
relativistic jets, provided that substantial deceleration can take
place on scales of a few kpc while allowing mildly relativistic
velocities to persist at 10-kpc scales. Environmental asymmetries
may affect the low-velocity edges of the jets, but it appears that
transverse velocity gradients need not always cause longitudinal
apparent magnetic field.

\section*{ACKNOWLEDGEMENTS}

We are grateful to Paddy Leahy for allowing us to use the
short-baseline L-band VLA data described in the text, and to the
referee, Dr R.\ Laing, for a number of comments which allowed us to
improve the clarity of the paper. MJH acknowledges a research
studentship from the UK Particle Physics and Astronomy Research
Council (PPARC), and thanks the Astrophysics group at the University
of Bristol for support during the final stages of this paper's
preparation. The National Radio Astronomy Observatory is operated by
Associated Universities Inc., under co-operative agreement with the
National Science Foundation. This project made use of Starlink
facilities.

%\clearpage
\begin{table}
\caption{VLA observations of \Ss{3C296}}
\label{obs}
\begin{center}
\begin{tabular}{llrlr}
&\multicolumn{2}{c}{8.4 GHz}&\multicolumn{2}{c}{1.5 GHz}\\
Conf.&Date&$t_{int}$&Date&$t_{int}$\\
&&(mins)&&(mins)\\
A&1995/07/24&120&1995/07/24&45\\
B&1995/11/27&120&1987/12/08$^a$&60\\
C&1994/11/11&55&1988/03/10$^a$&50\\
D&1995/03/06&30&1988/10/07$^a$&80\\
\end{tabular}
\end{center}
\parbox{\linewidth}{\medskip $^a$: data kindly supplied by J.P. Leahy.\\
$t_{int}$ denotes the total time spent on source at the specified VLA
configuration and frequency.}
\end{table}
\clearpage

\end{document}